\documentclass[prl,twocolumn,superscriptaddress,notitlepage
				]{revtex4-1}

\usepackage{epsfig}
\usepackage{graphicx}
\usepackage{color}
\usepackage{amsmath}
\usepackage{amssymb}
\usepackage{bm}

\newcommand{\e}{\varepsilon}

\newcommand{\w}{\omega}

\renewcommand{\aa}{\mathcal{A}}

\renewcommand{\Re}{\, {\rm Re} \,}
\renewcommand{\Im}{\, {\rm Im} \,}

\newcommand{\vc}[1]{{\bm #1}}
\newcommand{\tl}[1]{\tilde{#1}}

\newcommand{\comments}[1]{}

\newcommand{\co}{\hat{c}}
\newcommand{\ca}{\hat{c}^\dag}
\renewcommand{\ao}{\hat{a}}
\renewcommand{\aa}{\hat{a}^\dag}
\newcommand{\no}{\hat{n}}
\newcommand{\br}{{\bm r} }
\newcommand{\be}{{\bm e} }
\newcommand{\bq}{{\bm q} }
\newcommand{\bp}{{\bm p} }
\newcommand{\bk}{{\bm k} }
\newcommand{\bl}{{\bm l} }
\newcommand{\bQ}{{\bm Q} }
\newcommand{\bz}{{\bm 0} }
\newcommand{\ri}{{\mathrm i} }
\newcommand{\la}{\langle}
\newcommand{\ra}{\rangle}

\begin{document}

\title{Spontaneous time-reversal symmetry breaking for spinless fermions on a triangular lattice}

\author{O. Tieleman}
\affiliation{ICFO -- Institut de Ci\`{e}ncies Fot\`{o}niques, Parc Mediterrani de la Tecnologia, E-08860 Castelldefels, Spain}

\author{O. Dutta}
\affiliation{ICFO -- Institut de Ci\`{e}ncies Fot\`{o}niques, Parc Mediterrani de la Tecnologia, E-08860 Castelldefels, Spain}

\author{M. Lewenstein}
\affiliation{ICFO -- Institut de Ci\`{e}ncies Fot\`{o}niques, Parc Mediterrani de la Tecnologia, E-08860 Castelldefels, Spain}
\affiliation{ICREA -- Instituci{\'o} Catalana de Recerca i Estudis Avan\c{c}ats, Lluis Companys 23, E-08010 Barcelona, Spain}

\author{A. Eckardt}
\affiliation{ICFO -- Institut de Ci\`{e}ncies Fot\`{o}niques, Parc Mediterrani de la Tecnologia, E-08860 Castelldefels, Spain}
\affiliation{Max Planck Institute for the Physics of Complex Systems, Noethnitzer Str. 38, D-01187 Dresden, Germany.}

\date{\small\it \today}
\newcommand{\tred}[1]{{\color{red} \sout{#1}}}
\newcommand{\tblue}[1]{{\color{blue} #1}}

\begin{abstract}
As a minimal fermionic model with kinetic frustration, we study a system of 
spinless fermions in the lowest band of a triangular lattice with nearest-neighbor repulsion.
We find that the combination of interactions and kinetic 
frustration leads to spontaneous symmetry breaking in various ways. 
Time-reversal symmetry can be broken by two types of loop current patterns, a 
chiral one and one that breaks the translational lattice symmetry.  
Moreover, the translational symmetry can also be broken by a density wave 
forming a kagome pattern or by a Peierls-type trimerization 
characterized by enhanced correlations among the sites of certain triangular 
plaquettes (giving a plaquette-centered density wave). We map out the phase 
diagram as it results from leading-order Ginzburg-Landau mean-field theory.
Several experimental realizations of the type of system under study are
possible with ultracold atoms in optical lattices.
\end{abstract}

\maketitle

Geometric frustration in classical and quantum many-body systems is a source
of intriguing phenomena like extensive ground-state entropies, topological
order, and exotic emergent low-energy physics 
\cite{Wen04,Lhuillier05,AletEtAl06,MoessnerRamirez06,Sachdev08}. It is naturally 
encountered in systems of antiferromagnetically coupled localized magnetic
moments, arranged in a non-bipartite (e.g.\ triangular) lattice geometry that 
prohibits the favored antiparallel orientation between all pairs 
of neighboring moments. 
Recently, geometric frustration has also been induced
in the \emph{kinetics} of a system of ultracold bosonic atoms on a triangular
lattice by dynamically inverting the sign of the tunneling matrix 
elements \cite{eck10,str11}. For weak interaction the system 
shows spontaneous time-reversal (TR) symmetry breaking, whereas strong 
interactions are conjectured to lead to spin-liquid-like quantum disordered 
behaviour.

\begin{figure}[t]
\begin{center}
\includegraphics[width=.495\textwidth]{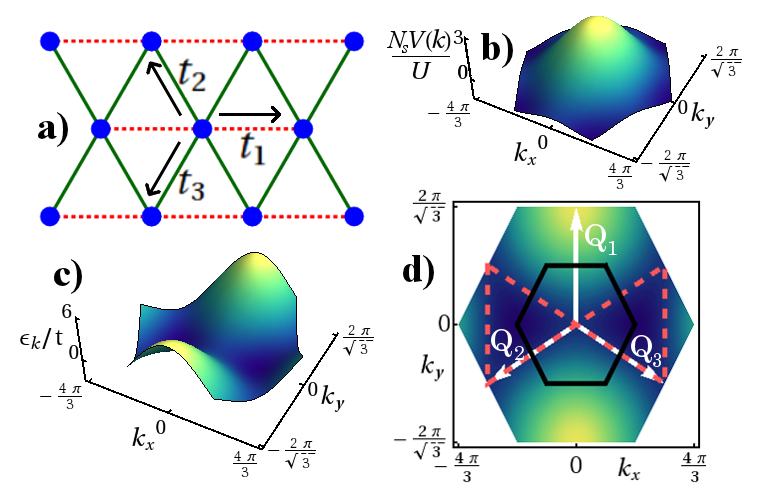}
\end{center}
\caption{(color online)
(a) Frustrated triangular lattice, with solid (dashed) lines representing 
postive (negative) hopping parameters $t_i$; the arrows indicate the vectors ${\bm e}_i$. 
(b) Momentum representation of the nearest neighbor interaction of 
strength $U$. 
(c) Single particle spectrum with two minima as a consequence of frustration.
(d) First Brillouin zone with nested Fermi surface (dashed `bowtie'-shaped 
line) and nesting vectors (white arrows); the black line indicates the reduced 
Brillouin zone for 4-sublattice translational symmetry breaking.}
\label{fig:ham}
\end{figure}

Here we investigate a minimal \emph{fermionic} model with kinetic 
frustration: spinless fermions on a triangular lattice, with nearest-neighbor interactions.
It can be realized, e.g., with ultracold dipolar atoms or molecules 
in an optical lattice \cite{lew12,lah09,blo08}. In this system kinetic frustration appears naturally as 
a consequence of Fermi statistics. 
Namely, for filling well above one half particle per site, the system is governed by 
the low-energy states of the fermionic holes whose kinetics is determined by 
the sign-inverted tunneling matrix elements. We find that (depending on the 
filling) two types of loop currents can spontaneously emerge at experimentally 
accessible temperatures (a chiral one and one that breaks translational 
symmetry). Thereby TR symmetry is broken neither as a consequence of coupling 
the kinetics to further degrees of freedom (such as spin or sublattice orbital 
freedom in the unit cell) nor in a process involving (quasi)long-range order 
in a continuous degree of freedom (i.e. Bose condensation). Typically at least 
one of these two themes is encountered when TR symmetry is broken in fermionic 
systems; examples range from ferromagnetic metals to Mott-insulators with spin 
or orbital order, but comprise also exotic states like chiral spin liquids 
(not featuring Bose condensation) \cite{WenEtAl89} or $p+ip$ superconductors 
(a condensate of pairs, not requiring spin) 
\cite{KohnLuttinger65,Volovik99,ReadGreen00,Ivanov01}.

We also observe that the nesting property of the Fermi surface 
near hole filling of $1/4$ gives rise to a rich spectrum of three different
types of instabilities that break the translational symmetry: One of them 
leads to spatially modulated currents (MC) and has been mentioned already. The 
other two give rise to a density wave (DW) and to a Peierls-type trimerization 
(PT, a plaquette-centered density wave).

\begin{figure}[t]
\begin{center}
\includegraphics[width=.495\textwidth]{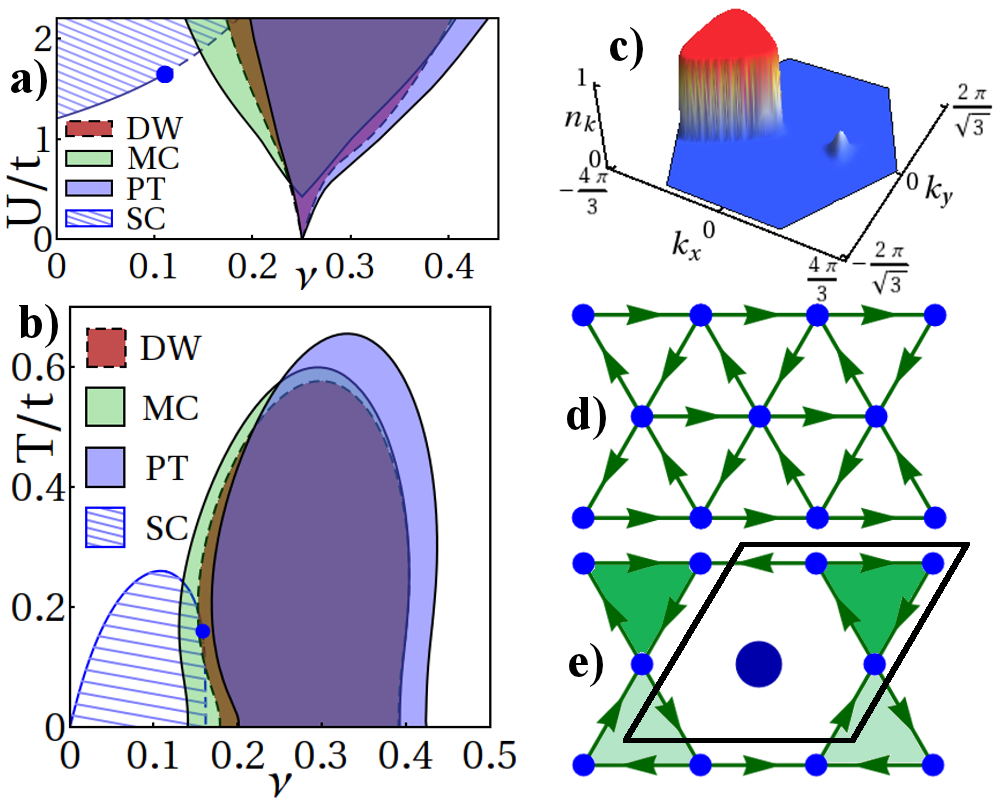}
\end{center}
\caption{(color online)
(a) and (b) Phase diagram resulting from low-order 
Landau expansion of the mean-field free energy, with respect to
interaction strength $U/t$, temperature $T/t$, and filling $\nu$. Solid 
(dashed) lines indicate second (first) order transitions; $T/t=0$ in a) and
$U/t=2$ in b).
(c) Momentum density distribution example for SC phase.
(d) Real-space pattern of staggered loop currents.
(e) Three types of translational symmetry breaking with four-site unit cell 
(black line): Modulated currents (MC, arrows) defining a kagome lattice 
embedded into the triangular lattice, a density-wave (DW, larger dot 
represents increased site occupation), and Peierls-type trimerization [PT, 
triangular plaquettes with increased (lowered) correlations are indicated by 
dark (light) green shading.].}
\label{fig:summary}
\end{figure}

\paragraph{Model} Our system is governed by the Hamiltonian
\begin{align}\label{eq:ham}
H = \sum_\br \sum_{\br'=\br\pm\be_i} \bigg(
		- t_i \ca_\br \co_{\br'}
		+\frac{U}{2} 
			\ca_\vc{r} \ca_\vc{r'} \co_\vc{r'} \co_\vc{r}
\bigg).
\end{align}
Here $\co_\vc{r}$ denotes the annihilation operator for a fermion at the 
lattice site located at $\br$, and the sum runs over all pairs of neighboring sites (connected by vectors $\pm\be_i$; see Fig.~\ref{fig:ham}a). Interactions between different sites can be 
achieved in various ways, e.g. by using dipolar atoms or molecules that are 
polarized perpendicular to the lattice plane 
\cite{gri05,lah09,tre11,lu12,ni10,wu12} or as a superexchange process between 
neighboring sites in a mixed Mott insulator of fermions and bosons 
\cite{kuk03,lew04,eck10a,SugawaEtAl11}. For simplicity, we have used isotropic nearest-neighbor interactions of strength $U>0$ \footnote{\label{fn:lri}In the case of dipolar interactions, the restriction to nearest-neighbor constitutes an approximation. However, we expect that including longer-range terms the mean-field theory would yield qualitatively similar results, since both momentum-space attraction and nesting (described below) would still be present.}.
The tunneling matrix elements 
read $t_1 = -t$ and $t_2 = t_3 =  t$ with $t>0$ (see 
Fig.~\ref{fig:ham}a), giving a $\pi$-flux through each 
plaquette. This sign configuration corresponds simply to studying holes 
instead of particles (which makes a difference in the non-bipartite triangular 
lattice) in a momentum shifted reference frame 
\footnote{\label{fn:ph}Starting from the standard (zero flux) configuration 
for particles, $t_i>0$, a particle-hole transformation $\co_\br\to\ca_\br$, 
implying $\no_\br=\ca_\br\co_\br\to1-\no_\br$, inverts all tunneling matrix 
elements, $t_i\to -t_i$, while it leaves the interaction term (up to a 
constant) unchanged. We find it convenient to apply another transformation 
$\co_\br\to\exp(\ri{\bq}\cdot\br)\co_\br$ [describing a shift in quasimomentum 
by $\bq=(0,2\pi/\sqrt{3})$] that inverts back $t_2$ and 
$t_3$.}. Another option is to control the sign of the $t_i$ via lattice shaking \cite{eck10,str11}. The system of $N$ 
particles on $N_{\rm s}$ sites (with filling $\nu=N/N_{\rm s}$) is characterized, moreover, by
the temperature $T$.

The tunneling coefficients are such that the kinetic energy cannot be 
minimized between all neighboring sites at the same time, since negative 
(positive) $t_i$ favor a single particle state with a relative phase 
of $\pi$ ($0$) between neighboring sites; the system is kinetically frustrated 
\cite{eck10}. As a consequence, the single-particle dispersion relation 
$\varepsilon(\bk)=-\sum_i2t_i\cos(\be_i\cdot\bk)$ possesses two minima (see 
Fig.~\ref{fig:ham}c). It is possible, however, to maximize the 
kinetic energy on every pair of neighboring sites, so the system is 
sensitive to the frustration for low filling only, when the Fermi surface 
explores low-energy states. 

The double-well structure of the dispersion 
relation supports various interaction-driven instabilities. One of 
them results from interaction favoring particles to be 
close by in (quasi)momentum: The potential energy between two fermions with 
sharp momenta $\bp$ and $\bp+\bq$ reads $V(\vc{0})-V(\bq)\equiv 
\tilde{V}(\bq)$, where the Fourier transform of the interaction
$V(\vc{q}) = \comments{(2N_{\rm s})^{\!-1} \sum_\vc{r} e^{- i \br \cdot \bq} 
V(\vc{r})=}N_{\rm s}^{-1}\sum_i U\cos(\be_i\cdot\bq)$ decays with
$|\bq|$ (see Fig.~\ref{fig:ham}b), and the relevant momentum-dependent 
second term (the exchange term) obtains a minus sign from Fermi statistics. 
For low filling this can, despite the kinetic energy cost, favor an
imbalanced occupation of the two wells (see Fig.~\ref{fig:summary}c) 
corresponding to staggered loop currents (SC) (see Fig.~\ref{fig:summary}d). 

Another source of instability appears near filling $\nu=1/4$ where 
both minima are filled up. Here the Fermi surface forms a bow-tie 
(Fig.~\ref{fig:ham}d, dashed line) and becomes approximately {nested}. Namely, 
$\e_\vc{k}-\e_F \approx \e_F- \e_{\bk+\bQ}$ with Fermi energy $\e_F$ and the
same nesting vector $\bQ$ for a significant fraction of momenta $\vc{k}$ near 
the Fermi surface, making the kinetic energy cost for spatial modulation 
described by $\bQ$ low \cite{RiceScott75} \footnote{The nesting is not exact because the condition only holds approximately, near the Fermi surface.}.
The nesting occurs for three vectors $\vc{Q}_j$ 
(Fig.~\ref{fig:ham}d, white arrows; $\bQ_j$ and $-\bQ_j$ are equivalent since 
$2\bQ_j$ is a reciprocal lattice vector). We find different instabilities that 
break translational symmetry (see Fig.~\ref{fig:summary}e), to be described 
below. 

The model (\ref{eq:ham}) with nearest-neighbor repulsion has recently also 
been studied in the strong coupling limit $U/t\gg1$, where for intermediate 
filling ($1/3\le\nu\le 2/3$) frustrated interactions lead to DW order 
\cite{hot06,hot06a,nis09}. In contrast, we are approaching the problem 
from the weak-coupling (mean-field) regime at low filling ($\nu\lesssim 1/4$) and find TR 
symmetry breaking as a consequence of frustrated kinetics. Fermions on the 
kagome lattice have been investigated in Refs.~\cite{obr10,kie12}.

\paragraph{Method} We use mean-field theory as a first approach (see 
Refs.~\cite{mik11,hal94} for the square lattice). 
In momentum representation, $\ao_\bk=N_{\rm s}^{-1/2}\sum_\br
e^{-i\bk\cdot\br}\co_\br$, the interaction consists of terms
$V(\bq)\aa_{\bk+\bq}\aa_{\bp-\bq}\ao_\bp\ao_\bk$ that we approximate like 
$\aa_\bk\aa_\bl\ao_\bp\ao_\bq\approx
\xi_{\bk\bq}\aa_\bl\ao_\bp +\xi_{\bl\bp}\aa_\bk\ao_\bq 
-\xi_{\bk\bq}\xi_{\bl\bp}
-\xi_{\bk\bp}\aa_\bl\ao_\bq -\xi_{\bl\bq}\aa_\bk\ao_\bq 
+\xi_{\bk\bp}\xi_{\bl\bq}$ while asking for self consistency
$\xi_{\bk\bp}=\la\aa_\bk\ao_\bp\ra$. Allowing for spatial modulations 
described by the nesting vectors, we retain averages of the form
$\la\aa_{\bk-\bQ_j}\ao_\bk\ra$ with $j=0,1,2,3$ where $\bQ_0\equiv\bz$.
Accordingly, we have single-particle correlations 
	\begin{equation}
	\la\ca_{\br+\be_i}\co_\br\ra=\sum_{j=0}^3
		e^{i\br\cdot\bQ_j}c_{ij}
	\end{equation} 
where $ c_{ij}= N_{\rm s}^{-1}\sum_{\vc{p}}
e^{-i(\bp-\bQ_j)\cdot\be_i}\la\aa_{\bp-\bQ_j}\ao_\bp\ra$ (summing over one 
Brillouin zone).
Defining $\be_0=\bz$, the local density $n_\br=\la\ca_\br\co_\br\ra$ is
related to the four averages $c_{0j}\equiv\rho_j$. The current from 
$\br$ to $\br+\be_i$, namely $J_i(\br)=2t_i\Im 
\la\ca_{\br+\be_i}\co_\br\ra/\hbar$, is described by 
$\Im c_{ij}\equiv I_{ij}$ ($i\ne0$, using the equivalence of $\bQ_j$ and $-\bQ_j$). 
We also define $\Re c_{i\ne0 j}\equiv R_{ij}$ for
the real (non-current-generating) part of the correlations 
$\la\ca_{\br+\be_i}\co_\br\ra$. The terms $\rho_0=N/N_{\rm s}$ and 
$R_{i0}$ are non-zero already without symmetry breaking. 

\begin{figure}[t]
\begin{center}
\includegraphics[width=.495\textwidth]{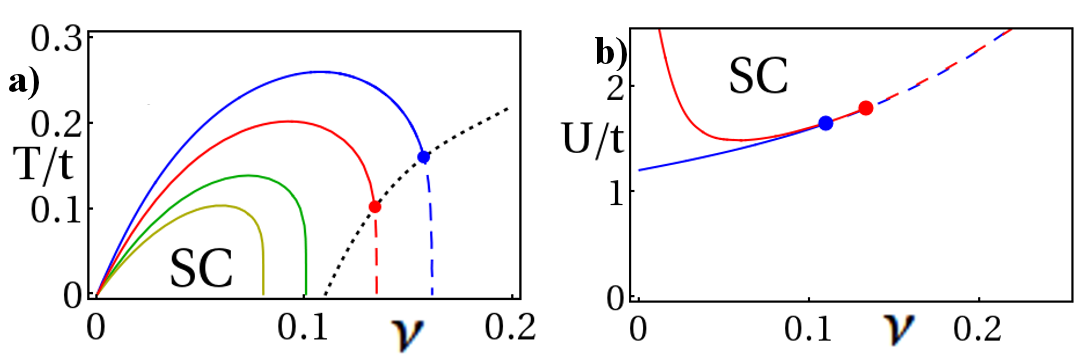}
\end{center}
\caption{(color online) 
Phase boundary of parameter region where staggered currents are 
expected (from low-order Landau expansion of the mean-field free energy)
Solid (dashed) lines indicate second (first) order transitions. (a) $U / t = 1.5, 1.6, 1.8, 2$, from small to large regions. (b) $T / t = 0$ (blue, crossing $y$-axis) and $T / t = 0.2$ (red, not crossing $y$-axis).}
\label{fig:stagcur}
\end{figure}

\paragraph{Staggered currents} Well below quarter filling, nesting is irrelevant and 
only terms with $j=0$ are important. The free energy reads
\begin{align}\label{eq:FSC}
F = - T \sum_{\vc{k}} \ln (1 + e^{- \w_\vc{k} / T}) 
- \sum_{\bp, \bq } \tl{V}(\bp-\bq) n_\vc{p} n_\vc{q},
\end{align}
with $n_\bk\equiv\la\aa_\bk\ao_\bk\ra$ and mean-field dispersion relation
$ \w_\vc{k} = 2\sum_{i = 1}^3\big[ ( t_i + U R_{i0}) \cos(\vc{k \cdot e}_i)
		+UI_{i0} \sin(\vc{k \cdot e}_i) \big]$. 
If $F$ is minimal for finite momentum imbalances 
$I_{i0}=N_{\rm s}^{-1}\sum_\bp\sin(\bp\cdot\be_i)n_\bp$, loop currents  will emerge.

In order to learn when this happens, we apply perturbation 
theory within the imaginary time formalism and Landau expand the free energy 
in powers of the $I_{i0}$:
	\begin{equation}\label{eq:GLSC}
	F\simeq F_0 +  {\bm I}_1^{\mathrm T} M {\bm I}_1 
	+ {\bm I}_2^{\mathrm T} K {\bm I}_2 + \mathcal{O}(I_{i0}^6).
	\end{equation}
Here ${\bm I}_n^{\mathrm T}=(I_{10}^n,I_{20}^n,I_{30}^n)$ 
(with power $n$ and transposition $\mathrm T$), $F_0$ is a constant, and  $M$ 
and $K$ are symmetric matrices. A second-order transition into a phase with 
finite ${\bm I}_1^{\mathrm T}=(-1,1,1)I$ occurs when $m_2$, the lowest eigenvalue of $M$, becomes negative, while higher-order contributions to $F$
remain positive.
The resulting phase boundaries are plotted as solid lines in 
Fig.~\ref{fig:stagcur}. 

If the quartic term, with matrix elements 
$K_{ij}=\delta_{ij}|\eta|+(1-\delta_{ij})\zeta$, becomes 
negative for $2\zeta<-|\eta|$ on the line defined by $m_2=0$, 
the phase transition becomes first-order (indicated by 
dashed lines in Fig.~\ref{fig:stagcur}) and the boundary shifts away 
from $m_2=0$, enlarging the symmetry broken region (neglected in 
Fig.~\ref{fig:stagcur}). Minimising the free energy (\ref{eq:FSC}) beyond the 
Landau expansion (\ref{eq:GLSC}) we find this shift to be small.

The current pattern in the symmetry-broken phase is defined by $J_i(\br)=J$, 
giving a staggered pattern of circular plaquette currents as shown in 
Fig.~\ref{fig:summary}d. The state is two-fold degenerate ($J = \pm |J|$) 
and chiral in the sense that both TR and lattice inversion transform one 
solution into the other \cite{WenEtAl89}. Experimentally this state can be 
identified by the momentum imbalance visible in time-of-flight absorption 
images.

The phase diagram (Fig.~\ref{fig:stagcur}) can be understood 
as follows. For lower densities, the slope of the single-particle spectrum at 
the Fermi surface is flatter (see Fig.~\ref{fig:ham}c), which explains why the 
critical interaction strength is reduced when the filling is lowered. 
If the filling is too low, however, a finite temperature $T$ may overcome the 
energy scale $U\nu$ of the interaction and destroy the currents altogether. 
Increasing the temperature counteracts the clustering effect from  
momentum-space attraction and makes symmetry breaking less favorable. 

For low filling $\nu<1/3$ no competing symmetry broken state is 
expected in the strong-coupling limit $U/t\gg1$ \cite{hot06,hot06a,nis09}. 
This makes it likely that the SC phase is not destroyed by quantum 
fluctuations beyond mean-field. We have performed exact diagonalizations for a 
small system (3 particles on 45 sites) at $T=0$, and find an increased 
admixture to the ground state of kinetic energy eigenstates with large $|I_{i0}|$ for 
$U/t\gtrsim1$, in support of the mean-field findings.

\paragraph{Spatial modulation} Near quarter filling, where nesting becomes relevant, we have to take into 
account the additional possibility of spatial modulation described by the 
nesting vectors. This leads to a triangular lattice with four-site unit cell 
(see Fig.~\ref{fig:summary}e), such that the mean-field Hamiltonian 
possesses four bands in a reduced Brillouin zone (see Fig.\ref{fig:ham}d). 
Broken translational symmetry is indicated by finite averages $\rho_j$, $R_{ij}$, or 
$I_{ij}$, with $j\ne0$, that open a gap between the lowest bands \footnote{The 
gap opens along the $\nu=1/4$-Fermi surface, forming a star-shaped structure 
in the reduced Brillouin zone.}. 
Expanding the free energy also with respect to those averages, in the 
leading (quadratic) order we find 
	\begin{equation}\label{eq:Fmod}
	F_\text{mod}^{(2)} =
	\sum_{j=1}^3\Big[ 
		{\bm D}_j^{\mathrm T}A {\bm D}_j
		+{\bm R}_j^{\mathrm T}B {\bm R}_j
		+{\bm J}_j^{\mathrm T}C{\bm J}_j
		+\lambda I_{jj}^2
		\Big]	
	\end{equation}
with vectors ${\bm D}_j^{\mathrm T}=(\rho_j,R_{jj})$, 
${\bm R}_j^\mathrm{T}=(R_{kj},R_{lj})$ and ${\bm J}_j^{\mathrm 
T}=(I_{k,j},I_{l,j})$ such that $j\to k\to l$ under 
cyclic permutation, symmetric $2\times2$ matrices $A$, $B$ and $C$, and 
coefficient $\lambda>0$. Thus $I_{jj}=0$ is favored. The remaining averages 
are coupled pairwise, with different nesting directions $j$ 
uncoupled in quadratic order.

\begin{figure}[t]
\begin{center}
\includegraphics[width=.495\textwidth]{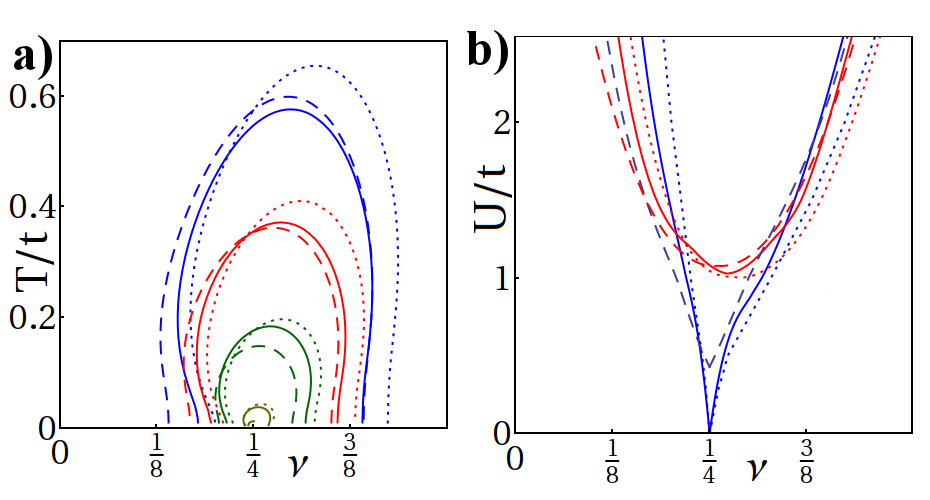}
\end{center}
\caption{(color online) 
Different types of instabilities that break the translational lattice symmetry
are found inside the plotted boundaries. Solid, dotted, and dashed lines stand
for instabilities leading to density-wave order, Peierls-type trimerization  
(i.e.\ plaquette-centered density-wave order), and a pattern of modulated 
currents, respectively. The transition to the density-wave order has to be
first order. (a) $U / t = 0.5$ (inner curve), $1, 1.5, 2$ (outer curves), (b) 
$T / t = 0$ (lower curves) and $0.1$ (upper curves).}
\label{fig:dw}
\end{figure}

A DW instability appears when the lowest eigenvalue of $A$ becomes 
negative (solid boundaries in Fig.~\ref{fig:dw}). In third order we find a 
term $\Gamma \rho_1\rho_2\rho_3$ that couples the density modulations along 
the three directions $j$. It favors the four-fold degenerate DW pattern shown 
in Fig.~\ref{fig:summary}e with the low-density sites defining a kagome 
lattice (and the coupling to $R_{jj}$ reducing the correlations among the 
low-density sites).
The third-order term also renders the transition first
order and shifts the phase boundary slightly (neglected in
Fig.~\ref{fig:dw}). Exact diagonalization studies with 4 particles on 16 sites
at $T=0$ confirm kagome DW patterns (with a slight anisotropy, probably a 
finite-size effect).

If the lowest eigenvalue of $B$ turns negative
(Fig.~\ref{fig:dw}, dotted lines), an instability towards a Peierls-type
modulation of nearest-neighbor correlations is found: 
The correlations $\la\ca_{\br'}\co_\br\ra$ on the dark (light) shaded 
triangular plaquettes in Fig.~\ref{fig:summary}e are increased (lowered); or 
vice versa. This eight-fold degenerate Peierls trimerization breaks also the 
inversion symmetry of the lattice and corresponds to a plaquette-centered DW. 

Finally, if $C$ acquires a negative eigenvalue (Fig.~\ref{fig:dw}, dashed 
lines) an instability towards the formation of MC is indicated. The
current carrying bonds define a kagome lattice with loop currents of equal 
orientation around both types of triangular plaquettes ($\bigtriangleup$ and 
$\bigtriangledown$) and of opposite orientation around the hexagonal 
plaquettes (see Fig.~\ref{fig:summary}e), such that the finite plaquette 
fluxes of the mean-field Hamiltonian vanish when averaged over the unit cell. 
This pattern breaks both TR and translational symmetry and is eight-fold 
degenerate.

The structure of the stability boundaries shown in Fig.~\ref{fig:dw} explains itself in an intuitive way. According to the nesting condition,
interaction-induced spontaneous translational symmetry breaking occurs for
filling near $1/4$, in a finite interval of filling factors
that grows with increasing interactions. An interesting effect is that for a
given interaction strength $U/t$ this interval has its maximum extent at a
non-zero temperature. The growth of the symmetry broken parameter region when
a small temperature is switched on, might be explained by the fact that
smoothening the Fermi edge increases the number of participating momentum
pairs for which the nesting condition is approximately fulfilled. However, for
large temperatures the order is eventually destroyed.  

As visible in Figure~\ref{fig:dw} there is a large region in parameter space, 
where according to second Landau order all three nesting-driven types of 
symmetry breaking instabilities appear. As long as the spontaneous kagome 
patterns defined by each of them match (as depicted in 
Fig.~\ref{fig:summary}e), they can in principle co-exist; their structures 
do not mutually exclude each other. This suggests that 
(especially as long as the order parameters are small such that low-order Landau 
terms dominate) the system might take advantage of
realizing two or even all three of them at the same time, leading to a
degeneracy of up to sixteen ($4 \times$ spatial, $2 \times$ time-reversal, and $2 \times$ inversion symmetry). However, generally it will depend on the 
energetics described by higher-order Landau terms that contain the coupling
between the different orders, whether mean-field theory
predicts such co-existences. Shifting and merging of the phase boundaries
due to higher-order terms can be expected.
Beyond mean-field theory this question can be
answered, for example, by an experiment with ultracold atoms.

All three translational symmetry breaking orders (DW, MC, and PT) open up a
gap between the lowest two bands of the mean-field dispersion relation in the
reduced Brillouin zone. However, the momentum-space structure of the coupling
that induces the gap depends on the type of order. Thus the three orders can
be distinguished by the momentum distribution, which is accessible through  
time-of-flight absorption imaging. The orders involving loop currents could, moreover, be observed using a quench in the tunnelling matrix elements 
\cite{KilliParamekanti12,kil12}. Exactly at quarter (hole) filling, an insulating phase is expected due to the gap, which should show up as a plateau (at particle filling $3/4$) in the density profile in a shallow trap.

We summarize our findings in Fig.~\ref{fig:summary} with subfigures (a) 
and (b) showing the combined phase diagram. For sufficiently large interactions
there is a small parameter region where both the SC and the MC overlap. We 
expect both types of order to repel each other and it appears likely that 
higher order terms will remove this overlap region and predict a direct 
transition between both phases.

We conclude that kinetic frustration, as it emerges naturally in a 
non-bipartite lattice potential at low filling of fermionic holes,
can give rise to rich physics without involving spin degrees of freedom. We 
find that for spinless fermions on a triangular lattice with nearest neighbor 
repulsion it gives rise to a number of interesting symmetry broken phases. 
They are characterized by chiral or modulated loop currents, Peierls-type 
trimerisation, and density-wave order. Our findings can be probed 
experimentally with ultracold fermionic atoms or molecules in optical lattices 
at accessible temperatures. Such an experiment would complement the recent 
observation of time-reversal symmetry breaking as a consequence of kinetic 
frustration in a system of bosonic atoms \cite{str11}. Many fascinating questions related to the findings discussed here can be investigated, including the role of quantum fluctuations beyond mean field, and such natural generalizations as the inclusion of a second spin species and on-site interactions. We hope to inspire such research with this paper.

\section{Acknowledgements}
OT thanks Achilleas Lazarides for helpful discussions and
acknowledges funding from the Netherlands Organization for
Scientific Research (NWO). Furthermore, we acknowledge funding
from AAII-Hubbard, EU STREP NAMEQUAM, ERC AdG QUAGATUA, Humboldt
Stiftung, Joachim Herz Stiftung, University of Hamburg and the foundation 
universidad.es.

\bibliography{Biblio}

\end{document}